\pdfoutput=1
\documentclass[aps,prb,twocolumn,showpacs,superscriptaddress,groupedaddress]{revtex4-1}

\usepackage{graphicx}  
\usepackage{dcolumn}   
\usepackage{amsmath}
\usepackage{amssymb}   
\usepackage{bm}        
\usepackage{physics}    
\usepackage{mhchem}    
\usepackage[subrefformat=parens]{subcaption} 

\let\originalleft\left
\let\originalright\right
\renewcommand{\left}{\mathopen{}\mathclose\bgroup\originalleft}
\renewcommand{\right}{\aftergroup\egroup\originalright}

\newcommand{\asi}{$a$-Si}
\newcommand{\csi}{$c$-Si}
\newcommand{\Tfive}{T_5}
\newcommand{\Tthree}{T_3}
\newcommand{\Tfour}{T_4}
\newcommand{\Tfoura}{T_{4\mathrm{a}}}

\begin{document}



\title{Analysis of single and composite structural defects in pure amorphous silicon: a first-principles study}
\author{Yoritaka Furukawa}
\author{Yu-ichiro Matsushita}
\affiliation{Department of Applied Physics, The University of Tokyo, Tokyo 113-8656, Japan}
\date{\today}

\begin{abstract}
The structural and electronic properties of amorphous silicon ({\asi}) are investigated by first-principles calculations based on the density-functional theory (DFT), focusing on the intrinsic structural defects.
By simulated melting and quenching of a crystalline silicon model through the Car-Parrinello molecular dynamics (CPMD), we generate several different {\asi} samples, in which three-fold ($\Tthree$), five-fold ($\Tfive$), and anomalous four-fold ($\Tfoura$) defects are contained.
Using the samples, we clarify how the disordered structure of {\asi} affects the characters of its density of states (DOS).
We subsequently study the properties of defect complexes found in the obtained samples, including one that comprises three $\Tfive$ defects, and we show the conditions for the defect complexes to be energetically stable.
Finally, we investigate the hydrogen passivation process of the $\Tfive$ defects in {\asi} and show that the hydrogenation of $\Tfive$ is an exothermic reaction and that the activation energy for a \ce{H2} molecule to passivate two $\Tfive$ sites is calculated to be 1.05 eV.
\end{abstract}

\pacs{N/A}
\maketitle

\section{Introduction \label{sec:introduction}}

Amorphous materials, which lack long-range structural orders but still keep short-range orders, have been investigated for almost a half century \cite{mott1971,brodsky1985,morigaki}.
An important problem in the physics of such disordered materials is the understanding of their structural characteristics that incorporates the short-range order in the disordered atomic network and of their influence on the electronic properties.
Despite the substantial progress made in the past, our understanding of how the structural characteristics affect the electronic properties of disordered materials is still incomplete.

Amorphous silicon (\asi) is an important example of such amorphous materials.
It shows attractive properties that are distinct from crystalline silicon (\csi), such as the high light absorption coefficient and the large bandgap.
Its low deposition temperature and low fabrication cost, as well as those physical properties, allow the industry to use this material for thin-film devices including solar cells and transistors.
Upon miniaturization of electronic devices, clarification of how the atom-scale structural characteristics influence its electronic properties of {\asi} is crucial also from a technological viewpoint.

Accurate determination of the local atomic structures of {\asi} is a prerequisite of any quantitative theoretical approach to its physical properties.
In the early days, continuous-random-network models \cite{polk1971, weaire1974} were used to consider the local structures of amorphous materials.
Then molecular dynamics (MD) \cite{ding,luedtke1988,ishimaru1997} or Monte Carlo \cite{wooten} techniques combined with empirical interatomic potentials \cite{stillinger,biswas,tersoff} are used to obtain the radial distribution functions of {\asi}.
However, the validity of the empirical potentials is always an issue and a ``try and error'' approach has been continued.

The MD approach based on the first principles of quantum theory gets rid of the problem of the interatomic potentials.
Car and Parrinello invented a scheme in which the electron-electron interaction is treated in the density-functional theory (DFT) \cite{hohenberg,kohn} and Hellmann-Feynman forces are used to track the dynamics of ions \cite{car1985}.
In this scheme, a fictitious mass of the wavefunction (Kohn-Sham orbital) is introduced to perform efficient first-principles MD (Car-Parrinello molecular dynamics, CPMD) simulations.
The CPMD scheme has been applied to {\asi} \cite{car1988, stich1991} and the obtained radial distributions up to the nearest neighbor distance agree with the experiments \cite{fortner1989, kugler1989, laaziri1999} satisfactorily.
Other Born-Oppenheimer MD simulations based on the DFT have been performed for {\asi} and the nature of the short-range order has been partly clarified \cite{drabold,lee1994,alvarez2002,morishita2009}.
These MD simulations show that while the Si atoms mostly form four-fold $\Tfour$ configurations, whose bond angles are around 110 deg, there are three kinds of structural defects in {\asi} as well:
The three-fold ($\Tthree$), the five-fold ($\Tfive$), and the anomalous four-fold ($\Tfoura$) sites \cite{car1988,stich1991,lee1994}.
The difference between the $\Tfour$ sites and the $\Tfoura$ sites is that the bond angles of the latter are strongly distorted than those of the former.
These structural defects may induce deep levels in the energy gap and be responsible for the conduction and valence band tails \cite{brodsky1985,morigaki,drabold1991}.
The deep levels are indeed observed by the electron paramagnetic resonance (EPR) measurements \cite{brodsky1985,morigaki,brodsky1979} and identified as either $\Tthree$ or $\Tfive$ configuration \cite{pantelides1986}.

A problem in the first-principles MD simulations described above is the high defect density appearing in the simulation cell:
The density is an order of which is higher by two orders of magnitude than the experimentally determined value \cite{brodsky1979} $10^{19}$ cm$^{-3}$.
In the CPMD simulations, {\asi} is prepared by heating {\csi} to melt and then quenching the obtained liquid.
The discrepancy in the defect density from the experimental situation is mainly due to the unrealistic quenching rate in the CPMD simulations, which is usually more than 100 K/ps.
In this work, we cool liquid silicon with the speed of $\sim 10$ K/ps, which is slower than those used to prepare {\asi} samples in the past, and show that defect-free structures can be certainly obtained by the CPMD.
Furthermore, using the generated samples,
we tackle several questions in {\asi} that have not been understood fully.

Theoretical efforts were made to clarify the relationships between the structural and the electronic properties of {\asi} \cite{joannopoulos1973,singh1981,nichols1988,dong1998},
and it was revealed that the disordered network of Si atoms is a key to understanding its electronic states.
In this paper, we make a quantitative analysis of the electronic properties of {\asi} and make a clearer explanation of how the geometric properties of {\asi} affects its density of states (DOS) in association with its atomic configurations.

One of the important things that characterize the structural property of {\asi} is the formation of the structural defects.
It is commonly assumed that the $\Tthree$ defects are prevalent {\asi}, and previous researches have paid less attention to the other two types of the defects.
Moreover, little is known about the complexes that these structural defects possibly compose.
Study of defect complexes is important in that they might help us to understand the how defects are spacially distributed or how likely they are to be stable in a particular configuration.
To give further insights into these problems, we perform calculations focusing on the defect complexes found in our {\asi} samples.

Another topic relevant to the defects is the effect of hydrogenation.
Defects in {\asi} are the origins of deep levels in the mid-gap, which contribute to lowering the mobility of {\asi} and thus to degrading its quality as a material for semiconductor devices.
Therefore, industrially fabricated {\asi} contains a large number of H atoms which passivate the defects in the structure with.
The deep levels are known to be made up of two electronic states: the dangling bonds and the floating bonds, the latter of which is a state derived from the $\Tfive$ defects \cite{biswas1989}.
While H atoms are believed to mostly passivate the $\Tthree$ defects, not much attention has been paid to the passivation of $\Tfive$ defects.
As some researchers have mentioned earlier \cite{pantelides1986,fornari1999}, however, the $\Tfive$ defects might be prevalent in {\asi} and play a major role in forming the deep levels.
We provide discussions on how likely the $\Tfive$ defects are to be passivated by H atoms from an energetic point of view.

The outline of this paper is as follows.
In Section \ref{sec:setup}, the calculation methods and the process of generating {\asi} samples are described.
In Section \ref{sec:structure-all}, the structural properties of the obtained samples are investigated.
In Section \ref{sec:electronicStates-dos} we analyze the correspondence between the DOS and the structural properties of {\asi}.
In Section \ref{sec:electronicStates-defects}, we take a closer look at the electronic states near the Fermi energy, particularly focusing on the defect levels.
In Section \ref{sec:defect}, the stabilities of the defects are studied.
In Section \ref{sec:hydrogenation}, hydrogenation effect of the $\Tfive$ sites are discussed.
A summary and conclusions are given in Section \ref{sec:conclusion}.

\section{Computational details \label{sec:setup}}

We have used our RSDFT (Real-Space Density-Functional Theory) package \cite{iwata2010,iwata2014,rsdft}, in which the Kohn-Sham equation based on the DFT \cite{hohenberg1964} is calculated under the real-space scheme \cite{chelikowsky1994}.
In the real-space scheme, discrete grid points are introduced in the real space, and the wavefunction is expanded on the mesh in the real space.
To simulate the structures of {\asi}, CPMD calculations have been done using RS-CPMD (Real-Space Car-Parrinello Molecular Dynamics) code, which is incorporated in RSDFT package.
In this work, we have set the mesh size as 0.41 \AA, which corresponds to the cutoff energy of 31.7 Ry.
We have used PBE exchange-correlation functional \cite{perdew1996} and norm-conserving pseudopotential in both static and dynamic calculations.
Brillouin zone (BZ) sampling has been done for the $\Gamma$ point.
We have confirmed that the total energy of the system {\asi} converges within 0.2 eV/cell with our calculational settings.

Amorphous samples have been obtained by melting and quenching a crystalline structure through the CPMD simulations.
As for the initial structure, we have prepared a $3 \times 3 \times 3$ supercell containing 54 Si atoms/cell.
The volume of the supercell has been fixed as 1.08 $\mathrm{[nm^3]}$, which is consistent with the experiments.
The time step in the simulations has been set to be 0.1 fs, and the temperature has been controlled by velocity scaling.

We have started heating the system from 500 K and increased the temperature with the heating rate of 125 K/ps until it reaches 1700 K.
At 1700K, we have heated for another 5.8 ps to sufficiently liquify the system.
We have subsequently cooled the system until the temperature has reached 500 K.
Here, we have employed four different cooling rates separately: 20.0, 16.7, 14.3, and 12.5 K/ps.
We note that these speeds are slower than what have ever been applied for the CPMD simulation of {\asi} reported by other groups.
Finally, we have relaxed the structures of the final step of the cooling to obtain the stable atomic configurations.
We refer to the samples obtained from each cooling rate 20.0, 16.7, 14.3, and 12.5 K/ps as {\asi}20, {\asi}16, {\asi}14, and {\asi}12, respectively.
After the relaxation, each sample has been heated at 300K for 2.0 ps to calculate the radial distribution and the angle distribution, both of which are time-averaged functions of the atomic positions.

\section{Structures of the obtained samples \label{sec:structure-all}}

\begin{figure}[tb]
 \begin{minipage}[b]{0.4\linewidth}
   \centering
   \includegraphics[width=30mm]{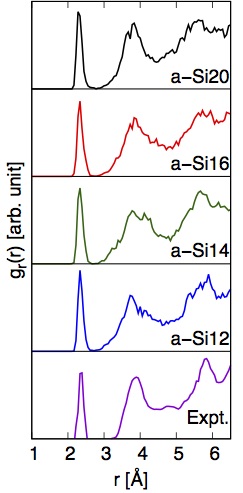}
   \subcaption{}
 \end{minipage}
 \begin{minipage}[b]{0.4\linewidth}
   \centering
   \includegraphics[width=30mm]{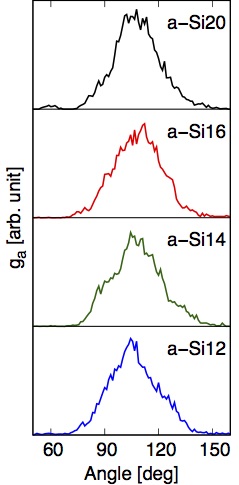}
   \subcaption{}
 \end{minipage}\\
 \begin{minipage}[b]{1.0\linewidth}
     \centering
     \includegraphics[width=60mm]{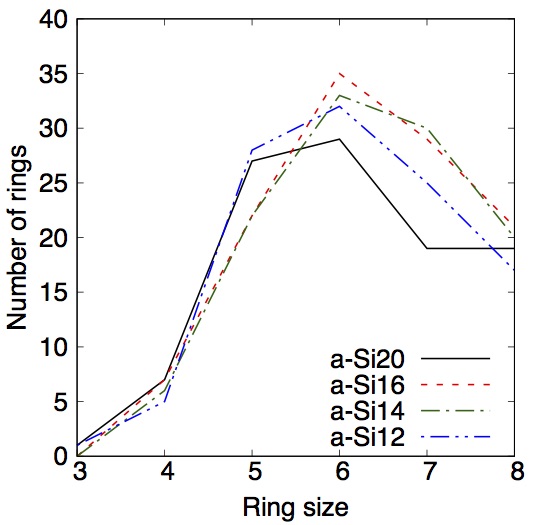}
     \subcaption{}
 \end{minipage}
   \caption{
   Structural properties of the obtained samples.
   (a) The radial distribution $g_r(r)$ with respect to the interatomic distance $r$ [\AA],
   (b) the bond angle distribution $g_a$ with respect to bond angle [deg],
   and (c) the number of $n$-membered rings in each obtained sample per supercell.
   Experimental data \cite{laaziri1999} is also shown in (a).
   \label{img:structure}}
\end{figure}

The radial distribution $g_r$ and the bond angle distribution $g_a$ of each sample are shown in Figure \ref{img:structure}.
All the four calculated plots of $g_r$ are in good agreement with the experimental result \cite{laaziri1999}.
The sharp peaks are located at 2.31, 2.33, 2.32, and 2.36 \AA~ for {\asi}20, {\asi}16, {\asi}14, and {\asi}12, respectively.
Considering the experimentally obtained first-neighbor distance in {\csi}, 2.35 \AA, these indicate that the short-range order is preserved in every amorphous sample.
Furthermore, we can clearly observe the second and third peaks near $r =$ 3.8 and 5.8 \AA. These broad peaks reflect the deviation in the bond length and the bond angle.

Every plot of $g_a$ shown in Figure \ref{img:structure} (b) has a broad peak around 100 deg.
The peak position is close to the bond angle in {\csi}, 109.5 deg, which indicates that the majority of the Si atoms in the samples retain the nearly-tetrahedral bonds.

Figure \ref{img:structure} (c) shows the distribution of the rings composed of $n$ ($n=3,\dots,8$) Si atoms in each sample.
We find that dominant rings are those composed of five or six atoms in every sample.
We also notice that three-membered rings are found only in {\asi}20 and {\asi}12, which can be associated with the small peaks around 60 deg in Figure \ref{img:structure} (b).

\begin{figure}[tb]
   \begin{center}
       \includegraphics[width=40mm]{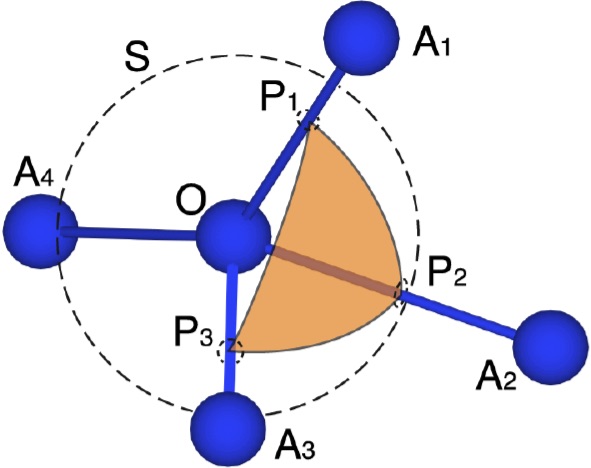}
       \caption{An example of a four-fold site, $O$, with four neighboring sites $A_1, A_2, A_3$, and $A_4$. $P_i ~ (i=1,2,3)$ is the point where $\protect\overrightarrow{OA_i}$ passes through the unit sphere $S$, represented by a dashed circle. The spherical triangle $P_1P_2P_3$ is colored by orange.
       \label{img:solidangle}}
   \end{center}
\end{figure}

\begin{figure}[tb]
 \begin{minipage}[b]{1.0\linewidth}
   \centering
   \includegraphics[width=60mm]{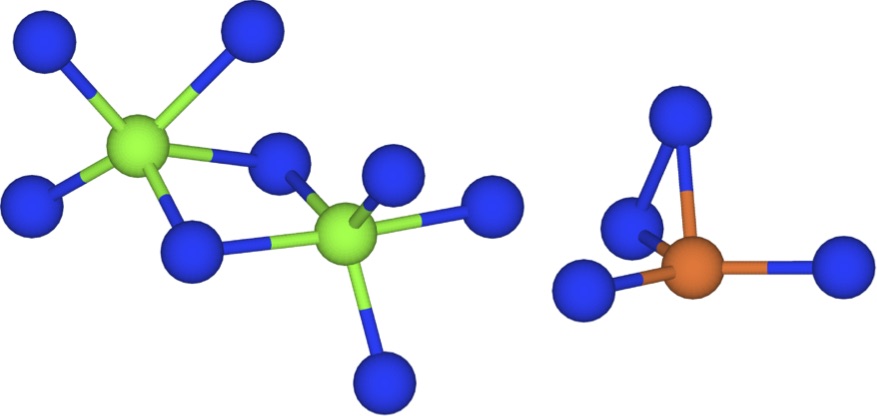}
   \subcaption{}
 \end{minipage}\\
 \begin{minipage}[b]{1.0\linewidth}
   \centering
   \includegraphics[width=35mm]{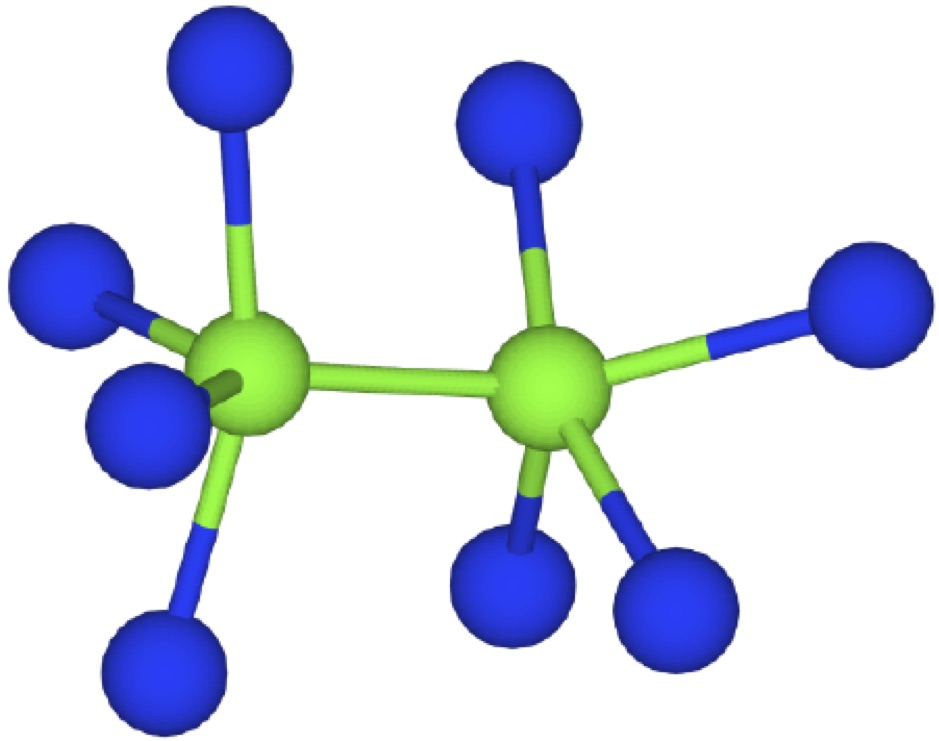}
   \subcaption{}
 \end{minipage}\\
 \begin{minipage}[b]{1.0\linewidth}
   \centering
   \includegraphics[width=60mm]{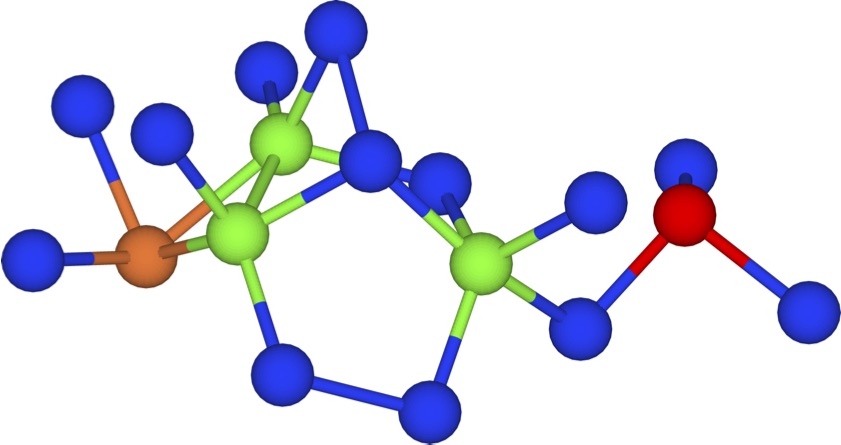}
   \subcaption{}
 \end{minipage}\\
 \begin{minipage}[b]{1.0\linewidth}
   \centering
   \includegraphics[width=25mm]{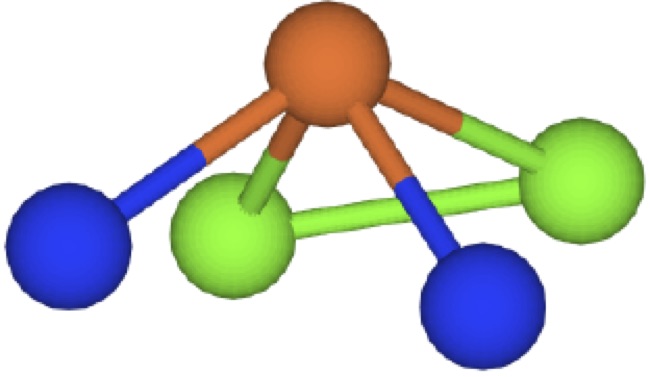}
   \subcaption{}
 \end{minipage}
 \caption{
 Observed defects in the three samples. $\Tfive$, $\Tthree$, $\Tfoura$, and normal $\Tfour$ sites are colored by green, red, orange, and blue, respectively.
 (a), (b), (c): Defects in {\asi}20, {\asi}16, and {\asi}12, respectively.
 Note that in {\asi}20, the $\Tfoura$ site is spacially separated from the $\Tfive$ sites.
 In (d), the $\Tfoura$ site in {\asi}12, which is highly distorted, is independently shown.
 \label{img:defects}}
\end{figure}

Defects found in the three samples are illustrated in Figure \ref{img:defects}.
In this paper, we define a $\Tfoura$ site as one which satisfies the following inequality
\begin{equation}
   \Omega = \sum_{(i,j,k)}\Omega_{ijk} < 4\pi.
   \label{eq:Tfoura}
\end{equation}
Here, $\Omega$ is the ``solid angle'' [sr] at the $\Tfoura$ site which can be described using Figure \ref{img:solidangle}.
Figure \ref{img:solidangle} schematically shows a $\Tfour$ site, named $O$, and the four neighboring sites $A_1, A_2, A_3$, and $A_4$. $P_i ~ (i=1,2,3)$ is the point where a vector $\overrightarrow{OA_i}$ passes through the unit sphere $S$, represented by the dashed circle.
$\Omega_{123}$ is the area of the spherical triangle $P_1P_2P_3$, colored by orange. $\Omega$ is the sum taken over for all the combinations of the four neighboring sites.
By this formulation, for those that retain nearly-tetrahedral bonds, such as the one illustrated in Figure \ref{img:solidangle}, $\Omega$ is equal to $4\pi$.
On the other hand, those that contain heavily distorted bonds, such as the one shown in orange in Figure \ref{img:defects}, satisfy Eq.\ (\ref{eq:Tfoura}).

The numbers of the defects ($N_{\Tfive}$, $N_{\Tthree}$, $N_{\Tfoura}$), the defect concentration $\rho_\mathrm{d}$, and the total energy $E_\mathrm{tot}$ of each obtained sample are listed in Table \ref{fig:structure}.
All the three kinds of the defects are incorporated into the calculation of $\rho_\mathrm{d}$.
Table \ref{fig:structure} reveals that {\asi}14 contains no defect, in contrast to the other three structures.
We also find that $\Tfive$ defects are prevalent in all the defect-containing samples.

$E_\mathrm{tot}$ of {\asi}14, the defect-free structure, is found to be the lowest of all. The highest is that of {\asi}12, which contains the largest number of the defects of all the samples. This implies that the existence of the defects increases the energy and thus decreases the energetic stability of the structure.
Furthermore, although it is expected that slower cooling rate makes the smaller amount of the defects, which is why we have introduced very slow cooling rates for this study, the defects are most abundant in {\asi}12, which has been produced with the slowest cooling speed.
We have found that this is of statistical occurrence as described below.

We have additionally generated ten different {\asi} samples employing the same cooling rate for {\asi}20.
The numbers of the defects and the total energy of each obtained sample are listed in Table \ref{fig:10samples}.
The variance of $E_\mathrm{tot}$ is 0.78 eV, and the difference between the maximum and the minimum energy is more than 3.0 eV.
Moreover, two samples are found to be free of defect, while the others contain 2 to 5 defect sites.
This way, we have demonstrated that the number of the defects and the total energy fluctuate even with the fixed cooling rates, and this suggests that the inconsistency between the total energy and the cooling speed in the former four samples can be explained as a statistical error.

\begin{table}[tb]
   \begin{ruledtabular}
       \caption{
       Properties of the obtained samples:
       The number of $\Tthree$, $\Tfive$, $\Tfoura$ sites $N_{\Tthree}$, $N_{\Tfive}$, $N_{\Tfoura}$ per supercell,
       the defect density $\rho_{\mathrm{d}} \mathrm{[cm^{-3}]}$,
       and the total energy $E_\mathrm{tot}$ [eV] per supercell with respect to that of {\asi}14.
       Here, the cutoff distance for first neighbors is set to be 2.7 \AA.
       }
       \begin{tabular}{ccccccc}
           Structure & $N_{\Tfive}$ & $N_{\Tthree}$ &$N_{\Tfoura}$ & $\rho_\mathrm{d}$ & $E_\mathrm{tot}$\\
           \hline
           {\asi}20 & 2 & 0 & 1 & $2.78\times10^{21} $ & +0.25 \\
           {\asi}16 & 2 & 0 & 0 & $1.85\times10^{21} $ & +0.19 \\
           {\asi}14 & 0 & 0 & 0 & $<9.25\times10^{20}$ &  0.00 \\
           {\asi}12 & 3 & 1 & 1 & $9.3\times10^{21}  $ & +1.14 \\
       \end{tabular}
   \label{fig:structure}
   \end{ruledtabular}
\end{table}

\begin{table}[tb]
   \begin{ruledtabular}
       \caption{Statistics of the properties of the 10 samples that have been obtained using the same generation procedure as that for {\asi}20:
       the numbers of three kinds of the defects $N_{\Tfive}$, $N_{\Tthree}$, and $N_{\Tfoura}$,
       the defect density $\rho_{\mathrm{d}} \mathrm{[cm^{-3}]}$.
       and the total energy $E_\mathrm{tot}$ [eV] with respect to that of the sample indexed as 8.
       }
       \begin{tabular}{ccccccc}
           Sample index & $N_{\Tfive}$ & $N_{\Tthree}$ &$N_{\Tfoura}$ & $\rho_\mathrm{d}$ & $E_\mathrm{tot}$\\
           \hline
           1        & 2    & 0   & 0   & $ 1.85\times10^{21}$ & +0.38 \\
           2        & 3    & 1   & 0   & $ 3.70\times10^{21}$ & +2.15 \\
           3        & 0    & 0   & 0   & $ 0                $ & +0.62 \\
           4        & 4    & 0   & 1   & $ 4.63\times10^{21}$ & +1.94 \\
           5        & 4    & 0   & 1   & $ 4.63\times10^{21}$ & +0.23 \\
           6        & 0    & 0   & 0   & $ 0                $ & +1.09 \\ 
           7        & 1    & 1   & 0   & $ 1.85\times10^{21}$ & +0.90 \\
           8        & 2    & 0   & 0   & $ 1.85\times10^{21}$ & (min) \\
           9        & 0    & 2   & 0   & $ 1.85\times10^{21}$ & +1.60 \\
           10       & 4    & 0   & 1   & $ 4.63\times10^{21}$ & +3.07 \\
           Average  & 2    & 0.4 & 0.3 & $ 2.50\times10^{21}$ & +1.12 \\
           Variance &      &     &     &                      &  0.78 \\
       \end{tabular}
   \label{fig:10samples}
   \end{ruledtabular}
\end{table}

\section{The density of states \label{sec:electronicStates-dos}}

\begin{figure}[tb]
   \begin{center}
       \includegraphics[width=70mm]{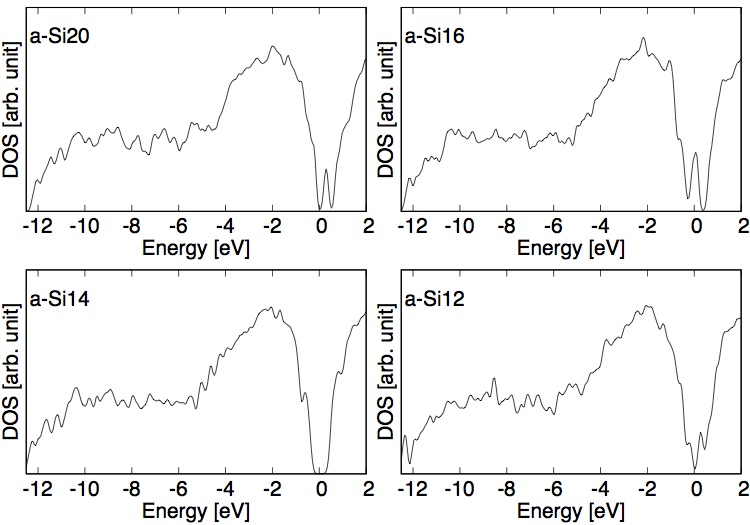}
       \caption{Density of states of the obtained samples. Zero energy is set to each Fermi energy.
        \label{img:dos-all}}
   \end{center}
\end{figure}

\begin{figure}[tb]
   \begin{center}
       \includegraphics[width=65mm]{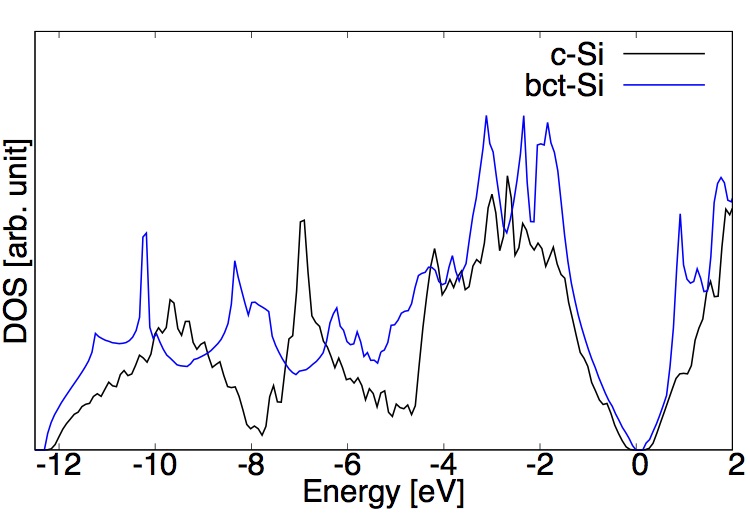}
       \caption{Density of states of {\csi} (black line) and bct-Si (blue line). Zero energy is set to each Fermi energy.
        \label{img:dos-csi-bct}}
   \end{center}
\end{figure}

Figure \ref{img:dos-all} shows the density of states (DOS) of each obtained sample.
Every plot has a strong peak at $\sim -2$ eV, which we shall call a ``high-energy'' peak, and a weaker, broader hump below $-6$ eV, which we shall call a ``low-energy'' hump.
These features make the DOS of {\asi} distinct from that of {\csi} shown in Figure \ref{img:dos-csi-bct}.
In the valence bands, the DOS of {\csi} possesses two low-energy peaks below -6 eV, with a valley in between, and a high-energy peak around -2 eV.
Joannnopoulos \textit{et al.} \cite{joannopoulos1973} made the following explanation for the difference of the shape of DOS between {\asi} and {\csi}.
They compared several polytypes of {\csi} and found that the DOS of ST-12 structure, which contains a five-membered ring in a unit cell, is similar to that of {\asi} in that the two low-energy peaks that are used to exist in {\csi} DOS are merged into one broad hump.
Then they concluded that the existence of odd-membered rings in the structure contributes to filling the gap between the two peaks at the lower energy.

Here, we give two perspectives on this argument.
Firstly, odd-membered rings are not necessary to fill the valley in the DOS.
In fact, a polytype of {\csi} called bct-Si \cite{fujimoto2008}, which contains only even-membered rings in the geometry, does not hold a valley in the low-energy part of its DOS; it is filled by another peak, as shown in Figure \ref{img:dos-csi-bct}.

\begin{figure}[tb]
   \begin{center}
       \includegraphics[width=60mm]{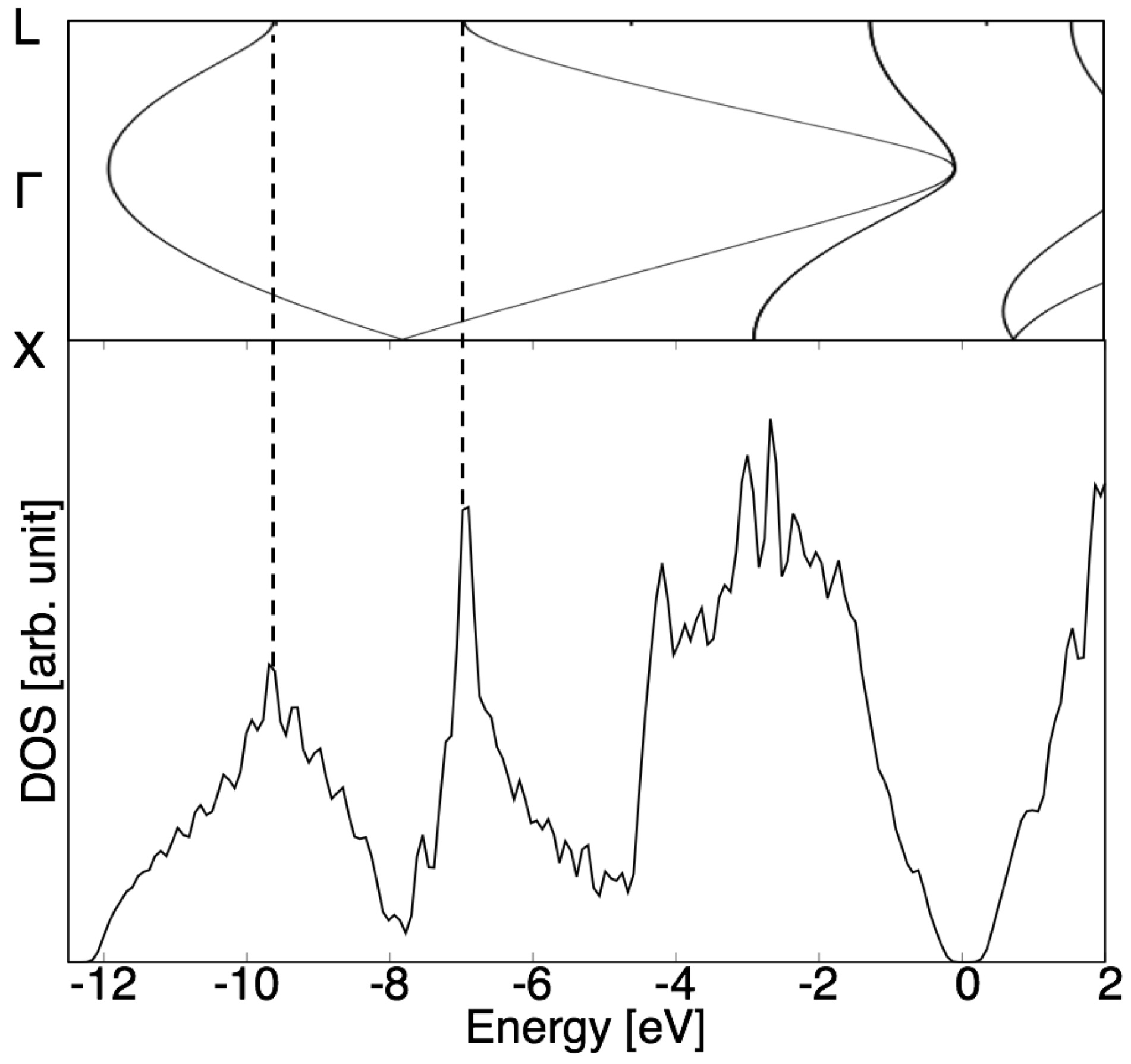}
       \caption{Comparison of the DOS and the band structure of {\csi}. Zero energy is set to each Fermi energy.
        \label{img:dos-csi-band}}
   \end{center}
\end{figure}

The second argument, which is more physically fundamental, is that the shape of DOS strongly depends on the symmetry of the system and thus its band structure.
We take {\csi} as an example here.
The correspondence between the DOS and the band diagram is illustrated in Figure \ref{img:dos-csi-band}.
We find that the position of the two low-energy peaks in the DOS of {\csi} correspond to the points in the band diagram where the energies of two bands flatten at the $L$ point, a symmetric point in the BZ.
We can understand that this is caused by Bragg reflection in the BZ.

The DOS satisfies the following equation:
\begin{equation}
   \mathrm{DOS}(E) \propto
   \int \frac{dS}{| \nabla_{\bm{k}} E(\bm{k}) |},
   \label{eq:dos}
\end{equation}
where $S$, $\bm{k}$, and $E$ represent the iso-energy surface in the reciprocal space, a reciprocal vector, and the energy.
Obviously from Eq.\ (\ref{eq:dos}), the DOS becomes larger with smaller $|\nabla_{\bm{k}} E(\bm{k})|$, and it should be peaky near $|\nabla_{\bm{k}} E(\bm{k})| \simeq 0$, which is the effect of Bragg reflection. The two peaks guided with dashed lines in Figure \ref{img:dos-csi-band} certainly demonstrate this.
One consequence of Eq.\ (\ref{eq:dos}) is that if the system loses its symmetry,
$| \nabla_{\bm{k}} E(\bm{k}) |$ no longer depends on $\bm{k}$
and thus the peaks in the DOS of the original structure will be vague or be totally lost.
This explanation is well fitted to what we have observed in the DOS of {\asi}:
In contrast to {\csi}, symmetry is lost in {\asi} and therefore Bragg reflection does not occur, ending up having no strict peak in its DOS.
This is one of the physical reasons that the DOS of {\asi} has a hump instead of peaks.

To summarize the point, the change of the DOS from {\csi} to {\asi} can be explained as the result of the loss of the symmetry of the system, rather than the emergence of the odd-membered rings.

This conclusion, however, does not explain how exactly the atomic configuration of {\asi} affects its DOS.
In the following paragraphs, we clarify this point by analyzing the DOS in detail in association with the geometry.

\begin{figure}[tb]
   \begin{center}
       \includegraphics[width=75mm]{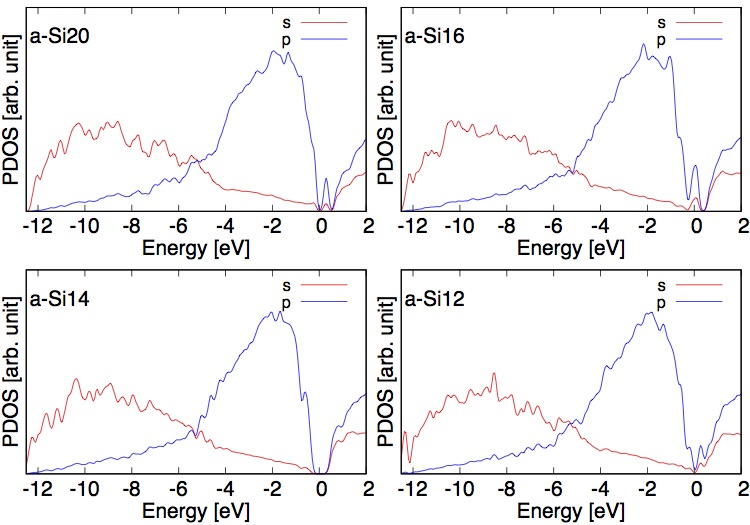}
       \caption{PDOS$_s$ of the obtained samples. Zero energy is set to each Fermi energy.
        \label{img:pdos}}
   \end{center}
\end{figure}

In Figure \ref{img:pdos}, we show partial density of states PDOS$_s$ and PDOS$_p$, which are DOS projected onto either $s$ or $p$ atomic orbitals, each defined as
\begin{equation}
   \mathrm{PDOS}_s(E) =
   \sum_i^{\substack{ \mathrm{all}\\\mathrm{atoms} }}
   \sum_j^{\substack{ \mathrm{occupied}\\\mathrm{states} }}
   |\braket{\phi^i_s}{\psi_j}|^2 \delta(E-E_j)
   \label{eq:pdos-s}
\end{equation}
\begin{equation}
   \mathrm{PDOS}_p(E) =
   \sum_i^{\substack{ \mathrm{all}\\\mathrm{atoms} }}
   \sum_j^{\substack{ \mathrm{occupied}\\\mathrm{states} }}
   |\braket{\phi^i_p}{\psi_j}|^2 \delta(E-E_j),
   \label{eq:pdos-p}
\end{equation}
where $\phi^i_s$, $\phi^i_p$ are the $s$ and the $p$ orbital of the $i$-th atom, and $\psi_j$ is the one-electron wavefunction of the $j$-th state.
For all the four samples, we observe that
PDOS$_s$ has a broad peak in the low-energy region, which implies that the low-energy hump of the DOS of {\asi} is mostly made up of the contribution of the $s$ orbitals.
One possible reason for PDOS$_s$ to have such a broad peak is that it has resulted from the variation in the Si-Si bond length, since the splitting of the bonding and antibonding energy becomes larger if two neighboring atoms get closer to each other, and vice versa.

To examine this idea, we further decompose the PDOS$_s$ into PDOS$^i_s$, which represents the contribution of atom $i$ to PDOS$_s$, and find a correspondence between the geometrical configuration of atom $i$ and the shape of PDOS$^i_s$.
$\mathrm{PDOS}^i_s$ is defined as
\begin{equation}
   \mathrm{PDOS}^i_s(E) =
   \sum_j^{\substack{ \mathrm{occupied}\\\mathrm{states} }}
   |\braket{\phi^i_s}{\psi_j}|^2 \delta(E-E_j).
\end{equation}
Compared with Eq.\ (\ref{eq:pdos-s}), this satisfies
$\mathrm{PDOS}_s(E) = \sum_i^{\mathrm{all~atoms}} \mathrm{PDOS}_s^i(E)$.

\begin{figure}[tb]
   \begin{center}
       \includegraphics[width=75mm]{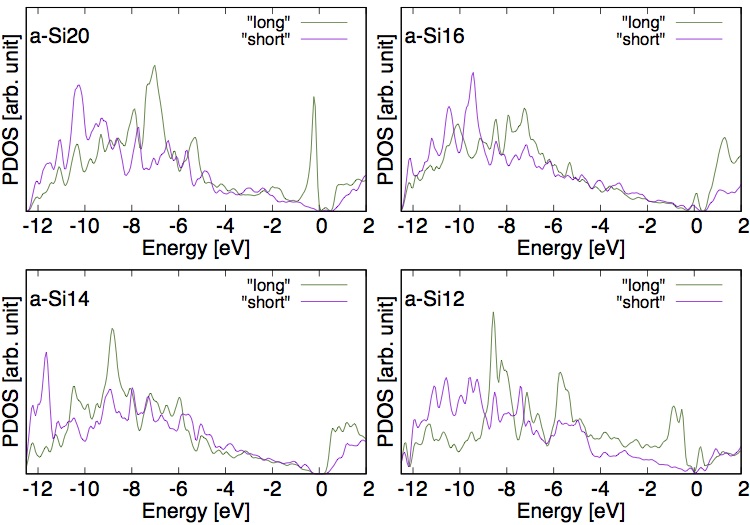}
       \caption{
       PDOS$^i_s$ of the obtained samples. Zero energy is set to each Fermi energy.
       The plot labeled as ``long'' (``short'') is for the atom whose bond length is the longest (shortest) of all the atoms in the system.
        \label{img:ldos}}
   \end{center}
\end{figure}

Figure \ref{img:ldos} shows PDOS$^i_s$ for each sample.
In each plot, the green (purple) curve is for the atom named ``long'' (``short''), whose bond length with its neighboring atoms is the longest (shortest), $r_\mathrm{max}$ ($r_\mathrm{min}$), in the sample.
The ratio $r_\mathrm{min}/r_\mathrm{max}$ of each sample is found to be within the range of 0.90 and 0.94.
To quantify the distribution of each PDOS$^i_s$, we define its ``center of mass'' $E_\mathrm{p}^i$ as
\begin{equation}
E_\mathrm{p}^i =
\frac{
   \displaystyle{
   \int_{E_\mathrm{bottom}}^{E_\mathrm{cut}}
   \mathrm{PDOS}^i_s(E)\cdot EdE
   }
}{
   \displaystyle{
       \int_{E_\mathrm{bottom}}^{E_\mathrm{cut}}
       \mathrm{PDOS}^i_s(E)\cdot dE
   }
}.
\end{equation}
Here, we have set $E_\mathrm{bottom}$ as -12.5 eV and $E_\mathrm{cut}$ as -4 eV to avoid the influence of the sharp peak near the Fermi energy that are originated from the defects.
Obtained values are presented in Table \ref{fig:ldos}.
We find that in every sample, $E_\mathrm{p}^\mathrm{short}$ is lower than $E_\mathrm{p}^\mathrm{long}$.
This implies that the peak position of PDOS$^i_s$ slides farther from the Fermi level with the shorter bond length of the atom.
These observations are consistent with the notion above and indicate a certain dependence on the bond length.
Therefore, we conclude that the variation of the Si-Si bond length in {\asi} broadens the peaks of PDOS$_s$, which is the sum of all PDOS$^i_s$, and this finally results in the emergence of the low-energy hump in the DOS.

\begin{table}[tb]
   \begin{ruledtabular}
       \caption{ Properties of PDOS$_s^i$:
       The bond length of atom ``short'' (``long''), $r_\mathrm{min}$ ($r_\mathrm{max}$),
       the ratio $r_\mathrm{min}/r_\mathrm{max}$,
       the ``center of mass'' of PDOS$^i_s$ of the atom ``short'' (``long'') $E_\mathrm{p}^{\mathrm{short}}$ ($E_\mathrm{p}^{\mathrm{long}}$).
       }
       \begin{tabular}{ccccccc}
           Sample & $r_\mathrm{min}$ [\AA] & $r_\mathrm{max}$ [\AA] & $r_\mathrm{min}/r_\mathrm{max}$ & $E_\mathrm{p}^{\mathrm{short}}$ [eV] & $E_\mathrm{p}^{\mathrm{long}}$ [eV] \\
           \hline
           {\asi}20 & 2.28 & 2.53 & 0.90 & -8.39 & -7.69 \\ 
           {\asi}16 & 2.28 & 2.49 & 0.92 & -8.43 & -8.10 \\ 
           {\asi}14 & 2.30 & 2.44 & 0.94 & -8.30 & -8.11 \\ 
           {\asi}12 & 2.30 & 2.50 & 0.92 & -8.40 & -7.51 \\ 
       \end{tabular}
   \label{fig:ldos}
   \end{ruledtabular}
\end{table}

\section{Electronic states \label{sec:electronicStates-defects}}

\begin{figure}[tb]
   \begin{center}
       \includegraphics[width=75mm]{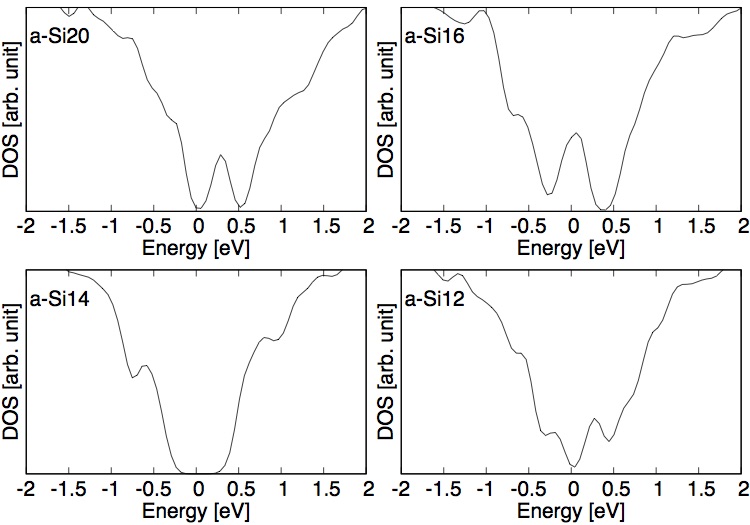}
       \caption{The DOS of the obtained samples near the Fermi energy.
        \label{img:dos-gap}}
   \end{center}
\end{figure}

\begin{figure}[tb]
 \begin{minipage}[b]{0.4\linewidth}
   \centering
   \includegraphics[width=32mm]{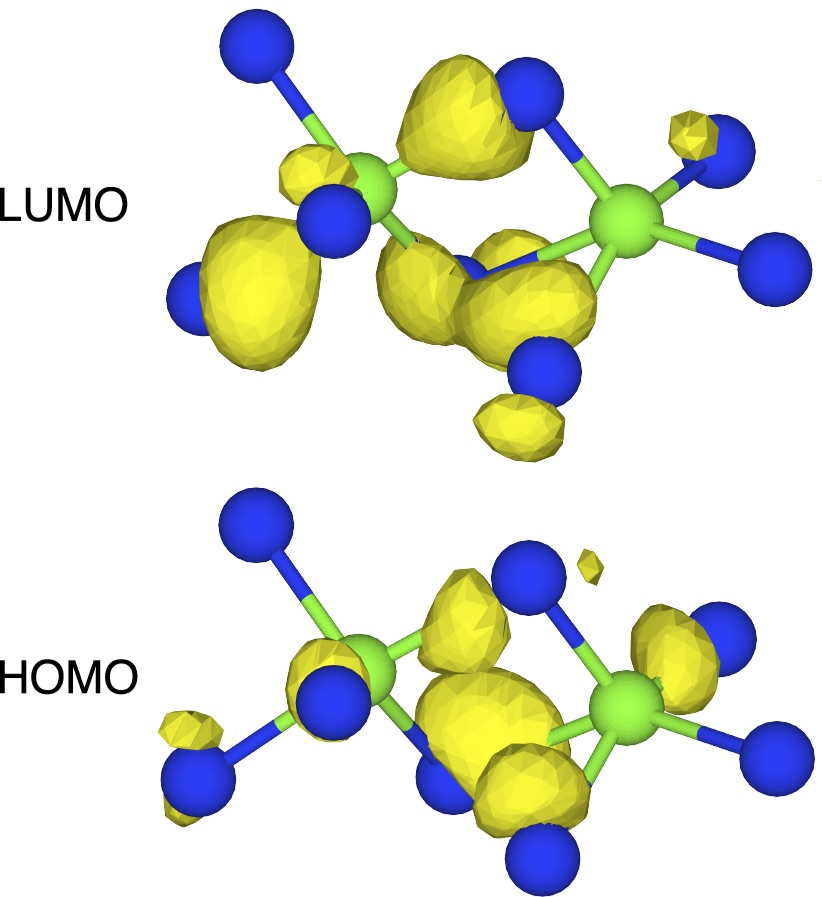}
   \subcaption{}
 \end{minipage}
 \begin{minipage}[b]{0.4\linewidth}
   \centering
   \includegraphics[width=29mm]{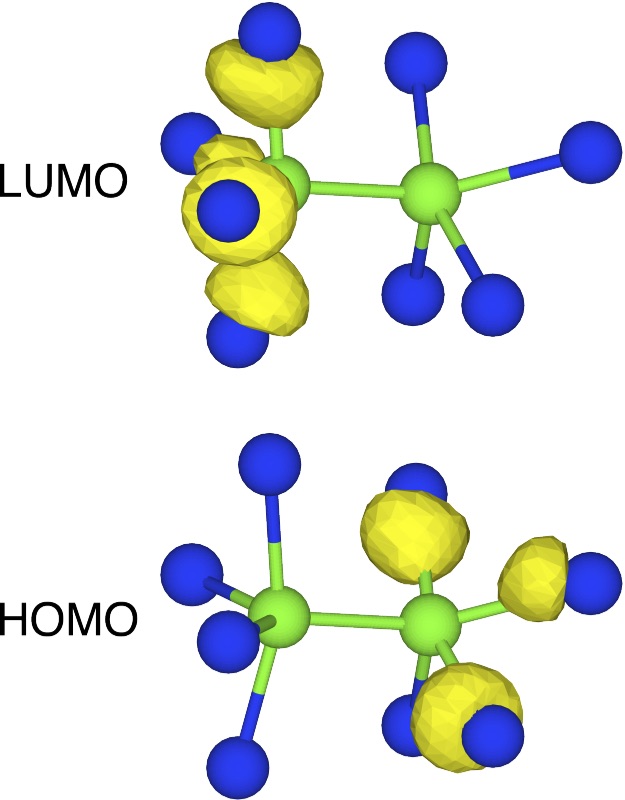}
   \subcaption{}
 \end{minipage}\\
 \begin{minipage}[b]{1.0\linewidth}
   \centering
   \includegraphics[width=67mm]{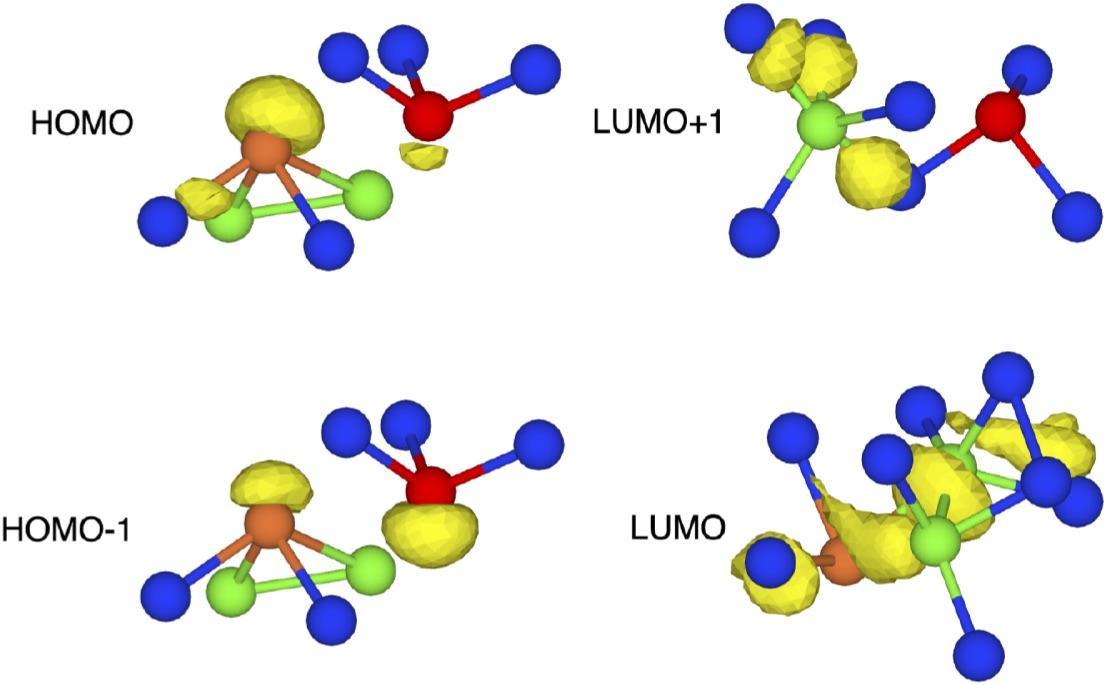}
   \subcaption{}
 \end{minipage}
 \caption{
 Electronic states near the Fermi energy for (a) {\asi}20, (b) {\asi}16, and (c) {\asi}12.
 The isovalue surface at 30 \% of the maximum is colored by yellow.
 Each $\Tfive$, $\Tthree$, $\Tfoura$, and normal $\Tfour$ site is colored by green, red, orange, and blue, respectively.
 \label{img:band-wf}}
\end{figure}

We take a look at the states near the Fermi energy, which are shown in Figure \ref{img:dos-gap}.
For the defect-containing structures ({\asi}20, {\asi}16, and {\asi}12) we observe small peaks, while a clear gap is seen for {\asi}14.
The mid-gap states of {\asi}20, {\asi}16, and {\asi}12 are understood by examining their electronic states shown in Figure \ref{img:band-wf}.
We immediately find that all the states are localized around the defects in each sample.

For {\asi}20 and {\asi}16, the highest occupied molecular orbital (HOMO) and the lowest unoccupied molecular orbital (LUMO) are found to be localized at the $\Tfive$ sites.
In contrast to the dangling bond found at the $\Tthree$ site and the $\Tfoura$ site in {\asi}12 (Figure \ref{img:dos-gap} (c)), the electronic states are rather delocalized around the $\Tfive$ defects, as pointed out in the past \cite{biswas1989}.
In {\asi}12, four localized states are found at $\Tthree$, $\Tfive$ and $\Tfoura$ sites.
We find that not all the defects are associated with the mid-gap states. The $\Tfoura$ site in {\asi}20, pictured in Figure \ref{img:defects} (a), makes no contribution in the gap states.
For {\asi}12, however, we find that the HOMO and the second highest occupied molecular orbital (HOMO-1) are due to the existence of the $\Tfoura$ site.
The important difference of the $\Tfoura$ site between {\asi}12 and {\asi}20 is that whereas in {\asi}20 it retains the nearly-tetrahedral bonds, in {\asi}12 it is heavily distorted, forming pyramid-like bonds with the neighboring atoms.

Using {\asi}14, we are capable of evaluating the bandgap of {\asi}.
From the DOS shown in Figure \ref{img:dos-gap}, calculated using PBE functional for the exchange-correlation energy, we obtain the bandgap of 0.96 eV.
To evaluate it more precisely, another calculation has been performed using HSE functional \cite{heyd2006}, which produces the bandgaps of covalent materials with higher precision \cite{matsushita2011}.
The same calculation has also been done for the {\csi} model.
Obtained bandgaps are listed in Table \ref{fig:bandgap}, along with experimental data \cite{forouhi1986, maley1987, jellison1996}.
The calculated bandgap of {\asi}14 is 1.41 eV, which is larger than that of {\csi} by 0.39.
The result is in good agreement with the experiments.

\begin{table}[tb]
   \begin{ruledtabular}
       \caption{Bandgaps obtained by PBE and HSE functionals.}
       \begin{tabular}{cccc}
                     &     & Bandgap & \\
           Structure & PBE & HSE  & Expt. \\
           \hline
           {\asi}14 & 0.96 & 1.41 & 0.95 - 1.4 \\
           {\csi}   & 0.67 & 1.02 & 1.1        \\
       \end{tabular}
   \label{fig:bandgap}
   \end{ruledtabular}
\end{table}

\section{Stability of complex defect structures \label{sec:defect}}

In {\asi}12, which contains four defect sites per supercell as illustrated in Figure \ref{img:defects} (c), we find that the two $\Tfive$ sites and one $\Tfoura$ site comprise a defect complex, forming a three-membered ring.
This result clearly shows the possibility of a defect complex consisting of two $\Tfive$ sites and one $\Tfoura$ site.

We can say that this defect complex is made up of three $\Tfive$ sites because of the following reason.
While the $\Tfoura$ atom in {\asi}12 is bonded with its four neighboring atoms, it holds a dangling bond at the same time, which means that it is capable of forming another chemical bond with another atom using the unbonded hand.
Therefore, we can regard the $\Tfoura$ site as a kind of $\Tfive$ site that lacks one neighboring atom and holds a dangling bond instead.
Using this notion, then, we can treat the defect complex in {\asi}12 as a triangle made up of three $\Tfive$ sites.
For the detailed analysis, we investigate the electronic properties and the stabilities of the defect complex observed in {\asi}12 by modeling ``defect-only'' structures as described below.

We consider three simple molecules composed of only $\Tfive$ sites: \ce{SiH5}, \ce{Si2H8}, and \ce{Si3H9}.
Each of them, illustrated in Figure \ref{img:sih5} - \ref{img:si3h9}, is a model of an isolated $\Tfive$ site, a neighboring $\Tfive$ sites, and a three-membered ring of $\Tfive$ sites, respectively.
Each Si atom is accompanied with five atoms of H and Si in which H atoms compensate for the missing neighboring bonds that exist in the actual structure of {\asi}.
By analyzing each model, we make the following discussions for the energetic stability of the $\Tfive$ defect complexes.
In the calculations below, to perform reliable calculations for systems containing H atoms, we have set the mesh spacing of the real space as 0.27 \AA, which corresponds to the cutoff energy of 72.5 Ry.

\subsection{\ce{SiH5}}
\begin{figure}[tb]
 \begin{minipage}[b]{0.4\linewidth}
   \centering
   \includegraphics[width=20mm]{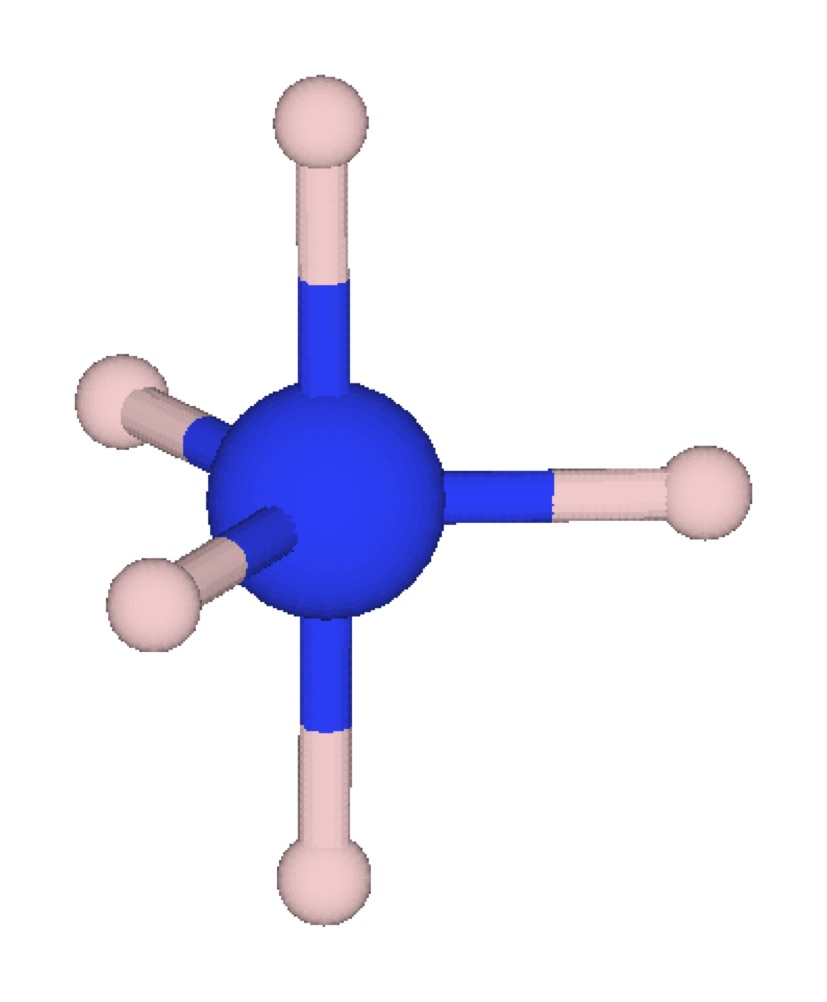}
   \subcaption{}
 \end{minipage}
 \begin{minipage}[b]{0.4\linewidth}
   \centering
   \includegraphics[width=20mm]{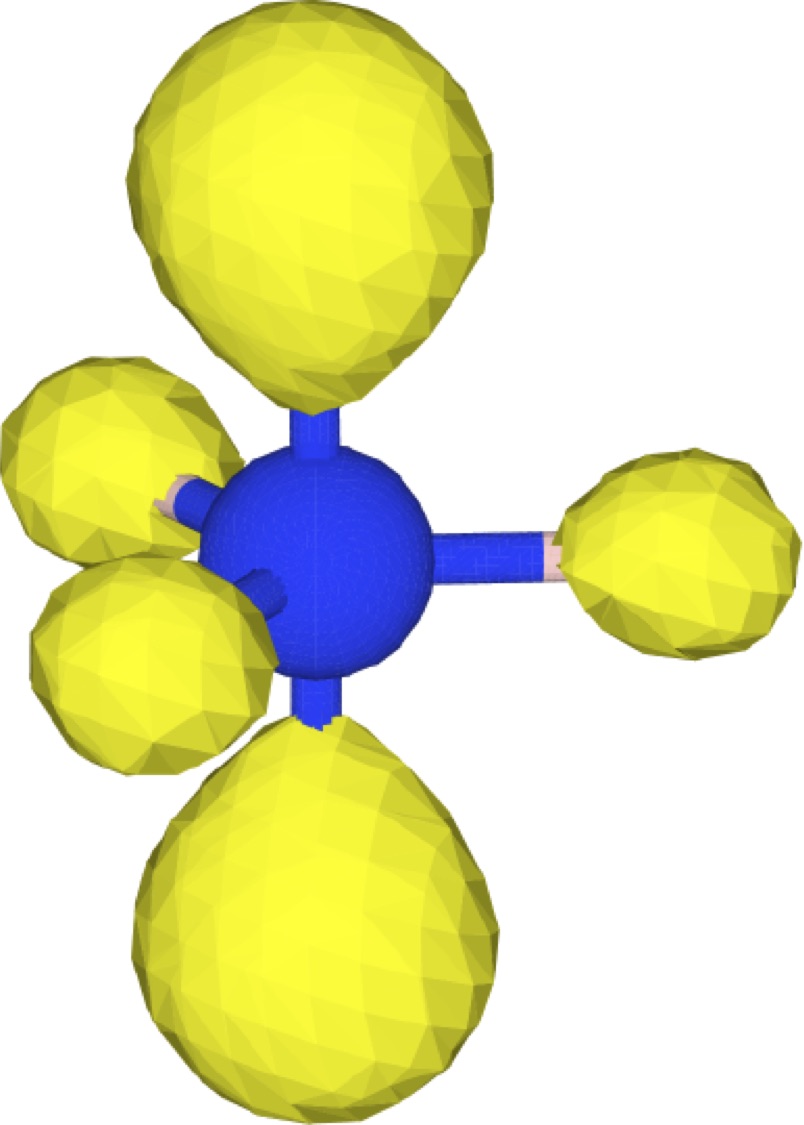}
   \subcaption{}
 \end{minipage}
 \caption{
 (a) The structure of SiH5 and (b) its wavefunction at HOMO, whose isovalue surface at 10 \% of the maximum is colored by yellow.
 The blue and pink sphere each represents a Si and an H atom
 (common for all the other figures).
 \label{img:sih5}}
\end{figure}

First, we clarify the electronic states of \ce{SiH5}.
The relaxed structure of \ce{SiH5} has a hexahedron-like shape, in which the five H atoms are covalently bonded to the Si atom in a stable manner, as shown in Figure \ref{img:sih5} (a).
The wavefunction of the HOMO presented in Figure \ref{img:sih5} (b) indicates that it is mostly the $s$ orbitals of the H atoms that comprises the HOMO and that it is a non-bonding orbital, which makes no contribution to the chemical bonds in \ce{SiH5}.
The result is not only consistent with the tight-binding analysis using parameters in Ref. \onlinecite{chadi1975},
but also with the character of the HOMO and LUMO wavefunctions of {\asi}20 and {\asi}16, which we have confirmed to have delocalized amplitudes around the $\Tfive$ sites, as shown in Figure \ref{img:band-wf} (a) and (b).

\subsection{\ce{Si2H8}}
\begin{figure}[tb]
 \begin{minipage}[b]{1.0\linewidth}
   \centering
   \includegraphics[width=35mm]{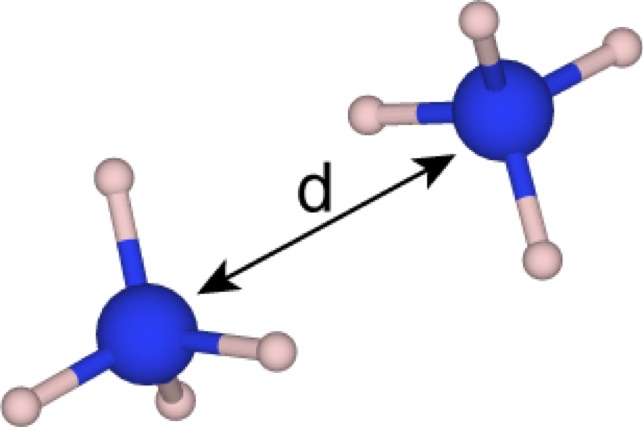}
   \subcaption{}
 \end{minipage}\\
 \begin{minipage}[b]{1.0\linewidth}
   \centering
   \includegraphics[width=60mm]{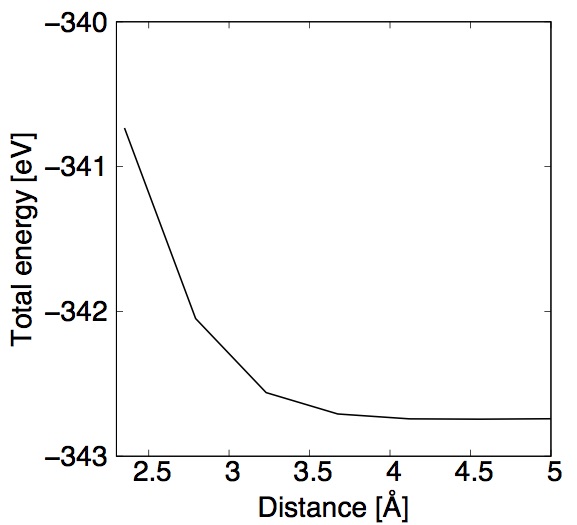}
   \subcaption{}
 \end{minipage}
 \caption{(a) Structure of \ce{Si2H8} and (b) its energy with respect to the distance between the two Si atoms $d$ indicated in (a).
 \label{img:si2h8}}
\end{figure}

Next, we show that the dimer \ce{Si2H8} is unstable.
We have calculated the dependence of the total energy of \ce{Si2H8} on the distance between the Si atoms, represented as $d$ in Figure \ref{img:si2h8} (a), by performing structural relaxations while fixing $d$ at several values.
The results are shown in Figure \ref{img:si2h8} (b), from which we immediately find that the total energy becomes smaller with longer $d$, meaning that neighboring two $\Tfive$ sites is energetically unstable.

\subsection{\ce{Si3H9}}
\begin{figure}[tb]
 \begin{minipage}[b]{0.4\linewidth}
   \centering
   \includegraphics[width=22mm]{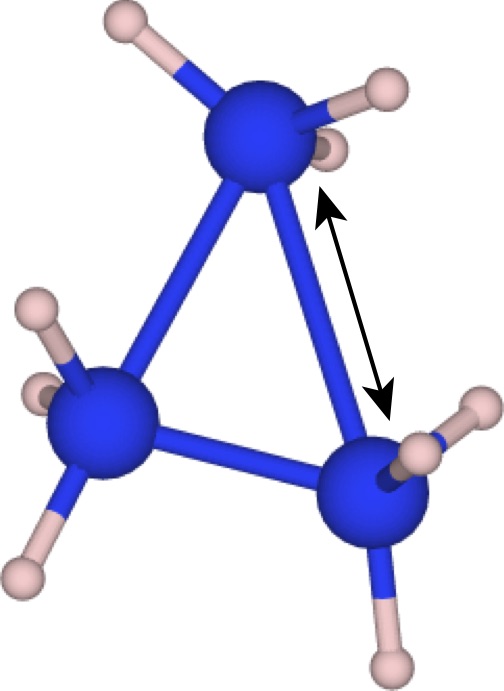}
   \subcaption{}
 \end{minipage}
 \begin{minipage}[b]{0.4\linewidth}
   \centering
   \includegraphics[width=25mm]{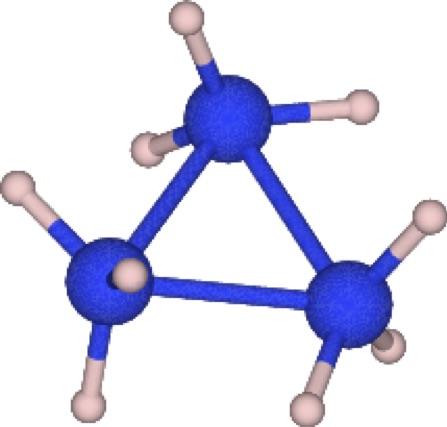}
   \subcaption{}
 \end{minipage}
 \caption{The structure of \ce{Si3H9} (a) in the neutral charge state and (b) in the positively charged state.
 \label{img:si3h9}}
\end{figure}

We finally show that \ce{Si3H9} is a stable configuration in the positively charged +1 state.
We have found that in the neutral charge, where the total number of the valence electrons is an odd number, the HOMO is an anti-bonding state occupied with an electron and that
it weakens the chemical bonds between the Si atoms.
Figure \ref{img:si3h9} (a) shows its relaxed structure.
Although it retains the triangular shape, the Si-Si bond indicated by an arrow is 3.25 {\AA}, which is much longer than what we have observed in {\asi} samples.

Another calculation has shown that by removing an electron from the structure, making it the positively charged state, the Si-Si bonds strengthens.
The relaxed structure of \ce{Si3H9} in the positively charged state is presented in Figure \ref{img:si3h9} (b).
The three Si atoms form a nearly-equilateral triangle, whose bond lengths and angles are almost 2.62 \AA and 60.0 deg, respectively.
These traits are close to that of the triangle in {\asi}12, where the atomic distances are within the range between 2.43 and 2.63 {\AA} and the bond angles between 54.7 and 62.9 deg.

Therefore, we conclude that \ce{Si3H9} is stable when it is positively charged and that the charge state of defect complex in {\asi}12 is thought to be close to that.

To summarize the point, by introducing three simplified models of the $\Tfive$ defects, we have found that the $\Tfive$ sites are likely to be formed in isolation from each other or as a positively charged trimer in a similar way as in {\asi}12, and that they are unlikely to form a dimer.

\section{The Effects of hydrogenating {\asi}}
\label{sec:hydrogenation}
In the previous sections we have observed that in our samples, $\Tfive$ is the most abundant of all the three kinds of the defects.
In this section, we make a quantitative analysis of the effect of passivating $\Tfive$ sites using H atoms, and examine whether this reaction energetically reasonable.

\begin{figure}[tb]
   \centering
   \includegraphics[width=75mm]{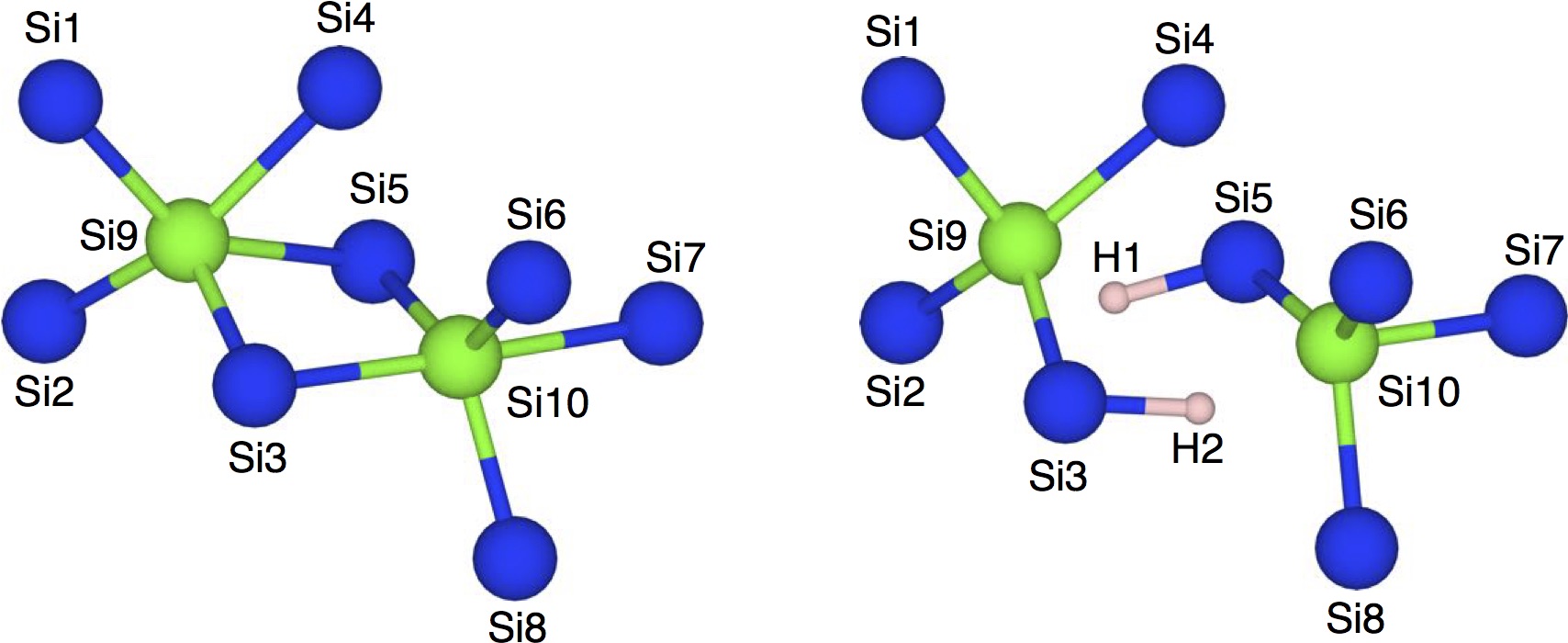}
   \caption{Atomic configurations around $\Tfive$ defects in {\asi}20 before (left) and after (right) hydrogenation. Every atom is labeled as either Si$n$ or H$n$, where $n$ is an integer. $\Tfive$ defects and H atoms are shown by green and pink spheres.
   \label{img:add2h-beforeafter}}
\end{figure}

We have chosen {\asi}20 as the target system for hydrogenation.
To generate Si-H bonds in it, we have manually inserted two H atoms near the $\Tfive$ sites and relaxed the structure. Figure \ref{img:add2h-beforeafter} shows the defects in {\asi}20 before (left) and after (right) hydrogenation. Si atoms are indexed by integers $n$, which we shall call each of them Si$n$.
We have found that two H atoms have broken into Si5-Si9 and Si3-Si10 bonds and that they have changed each $\Tfive$ site into a $\Tfour$ site, leaving the other bonds unchanged.
Considering that the lengths of both the Si5-Si9 and the Si3-Si10 bonds before hydrogenation, 2.68 and 2.45 {\AA} respectively, are longer than the first peak of the radial distribution, 2.31 {\AA}, these two Si-Si bonds are thought to be weak enough to allow the H atoms to break them and newly form Si-H bonds that are stronger than themselves.

\begin{figure}[tb]
   \centering
   \includegraphics[width=65mm]{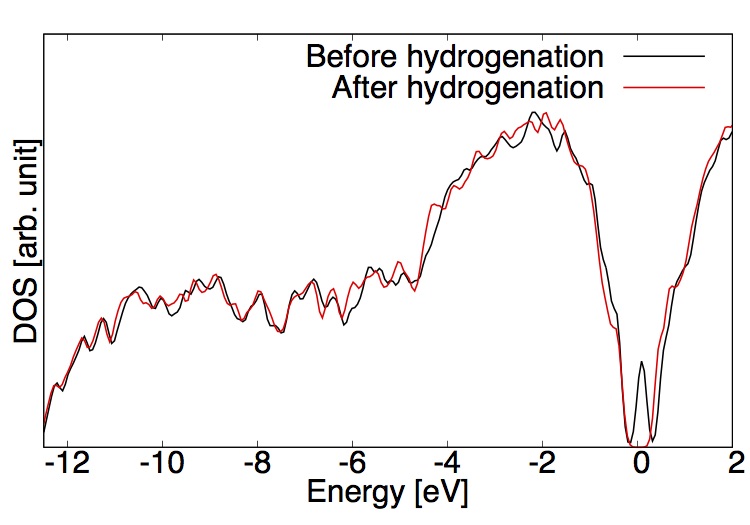}
   \caption{The DOS of {\asi}20 before (black line) and after (red line) hydrogenation.
   \label{img:add2h-dos}}
\end{figure}

The DOS before and after the hydrogenation are presented in Figure \ref{img:add2h-dos}.
We notice that the mid-gap peak, derived from the $\Tfive$ defects, has been completely removed after the hydrogenation, which is consistent with the computational result in the past study \cite{fornari1999}.

To clarity whether this type of transition is likely to occur from the energetic viewpoint, we have calculated the activation energy for a hydrogen molecule to passivate the $\Tfive$ sites by applying the Nudged Elastic Band (NEB) method \cite{mills1995} implemented in VASP code \cite{krasse1996}.
The NEB method provides a geometric pathway from the initial to the final state, assuring the continuity of each reaction step by imposing restrictions between different steps.

\begin{figure}[tb]
   \centering
   \includegraphics[width=75mm]{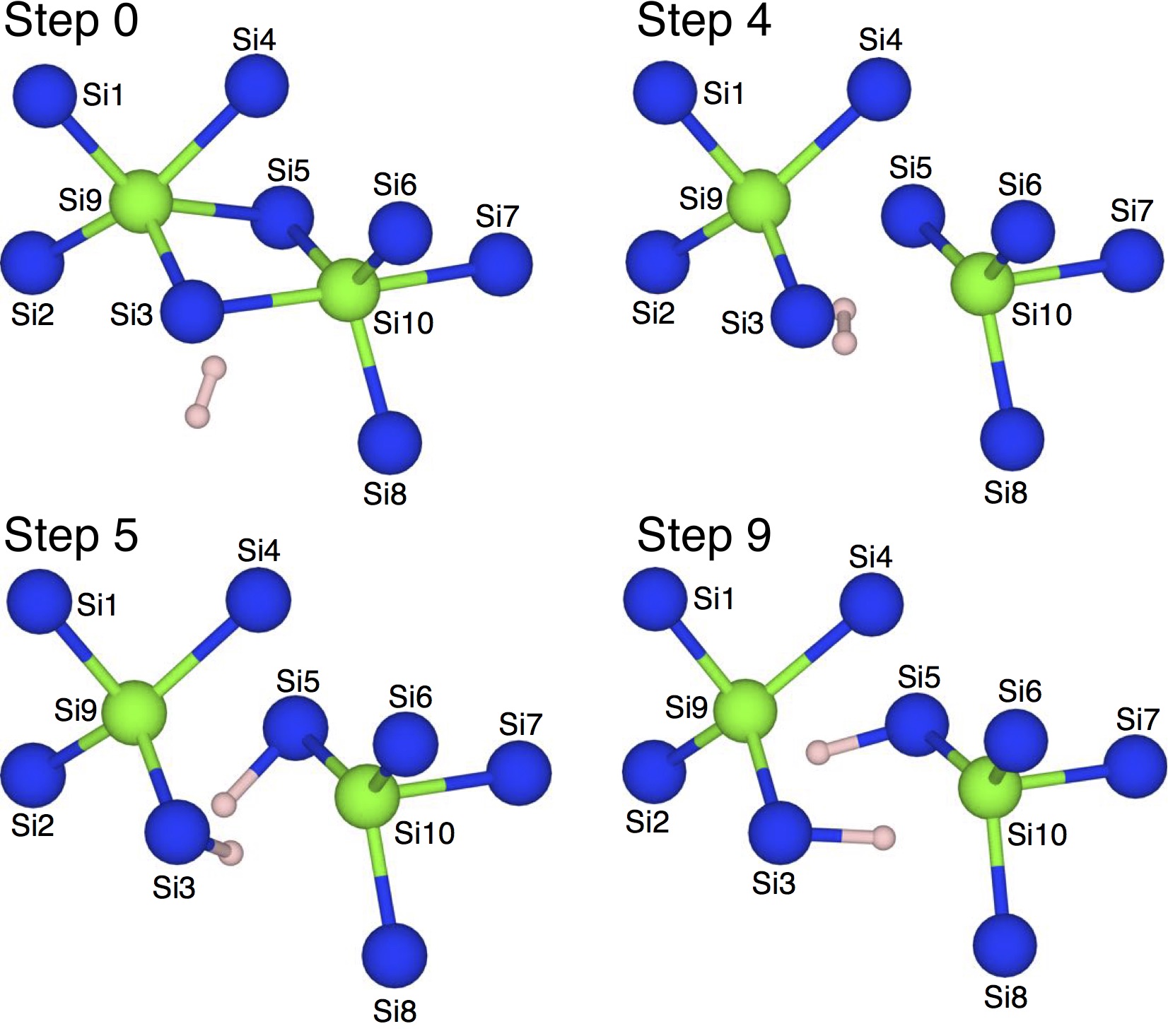}
   \caption{Transition from the initial state (step 0) to the final state (step 9) in the NEB method.
   \label{img:add2h-images}}
\end{figure}

\begin{figure}[tb]
   \centering
   \includegraphics[width=60mm]{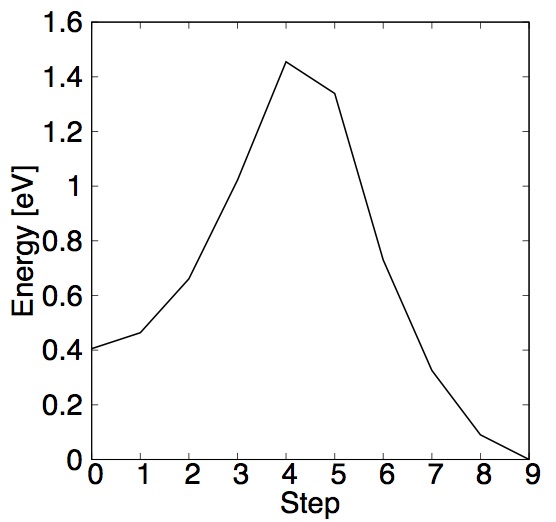}
   \caption{Energies at each NEB step in comparison with that of the final step.
   \label{img:add2h-energy}}
\end{figure}

We have prepared a structure which contains a \ce{H2} near the $\Tfive$ defects in {\asi}20, and set its optimized geometry as the initial step of the NEB calculation.
The final geometry has been chosen to be the one shown on the right in Figure \ref{img:add2h-beforeafter}.
Eight discrete steps have been imposed between the initial and final states.
The calculated total energy at each reaction step is shown in Figure \ref{img:add2h-energy} with respect to that of the final step.
From the comparison of the total energy of the initial and the final step, we confirm that this is an exothermic reaction of 0.41 eV, which implies that the \ce{H2} molecule certainly terminates the $\Tfive$ defects, leading to a stable atomic configuration.
The activation energy for a \ce{H2} to passivate $\Tfive$ sites is found to be 1.05 eV.
We find that the energy achieves the maximum at Step 4, whose geometry is shown in the top-right of Figure \ref{img:add2h-images}.
At Step 4, H atoms are still close to each other and retain the H-H bond with the distance of 0.91 \AA. The Si3-Si5 bond, in contrast, has been lost.
The results suggest that the energetic barrier 1.05 eV has been used to cut the bond between the $\Tfive$ defects.
In the next step, illustrated in the bottom-left of Figure \ref{img:add2h-images}, each H atom approaches the different Si atoms, namely Si3 and Si5, each of them forming a Si-H bond.
These are comparable to the diffusion energy of H in {\asi}:H $\sim1.5$ eV \cite{zellama1981}.
Therefore, the termination we have observed in our calculations is a reasonable one.

\section{Conclusion \label{sec:conclusion}}

In summary, we have performed first-principles calculations for {\asi}.
By CPMD simulation of melting and quenching of the crystalline structure,
we have obtained four samples of {\asi}, including one that contains no structural defect.
The radial distributions of all the samples have been in good agreement with the experiments,
and the angle distributions and the DOS have been consistent with the calculation result of the past studies.
We have found that three-fold ($\Tthree$), five-fold ($\Tfive$), and anomalous four-fold ($\Tfoura$) sites are generated in the samples and that the $\Tfive$ defects are more abundant than the other kinds of defects.
We have studied the origin of the transition of the shape of the DOS from {\csi} and {\asi}
and concluded that it is determined by the symmetry of the structure and is also certainly affected by the atomic configurations.
We have confirmed the emergence of the mid-gaps states in those that contain defects.
For the defect-free {\asi} sample, the bandgap has been found to be 1.41 eV from the HSE calculations, which is in good agreement with the experiments.
We have found a defect complex in a sample that consists of $\Tfive$ and $\Tfoura$ sites, and clarified that the $\Tfive$ sites can be stable in isolation from each other or by forming a trimer with a positive charge, and that the dimer of $\Tfive$ sites is energetically unstable.
Finally, the effect of hydrogenation on the $\Tfive$ site has been investigated by introducing H atoms.
We have found that two different $\Tfive$ defects have been hydrogenated through the exothermic reaction, and that the mid-gap states derived from the $\Tfive$ site disappear.
The activation energy for a \ce{H2} to passivate two $\Tfive$ defects has been determined to be 1.05 eV.

\begin{acknowledgements}
We would like to thank Professor Atsushi Oshiyama and Hirofumi Nishi for helpful discussions.
This work has been supported in part by Ministry of Education, Culture, Sports, Science and Technology. Computations have been performed mainly at the Supercomputer Center at the Institute for Solid State Physics, The University of Tokyo, The Research Center for Computational Science, National Institutes of Natural Sciences, and the Center for Computational Science, University of Tsukuba. This research partly used computational resources of the K computer provided by the RIKEN Advanced Institute for Computational Science through the HPCI System Research project (Project ID:hp160265). This work has been supported by JSPS Grant-in-Aid for Young Scientists (B) Grant Number 16K18075.
\end{acknowledgements}

\end{document}